\documentclass[final,5p,times]{elsarticle}
\usepackage{graphicx}
\usepackage{dcolumn}
\usepackage{bm}
\usepackage{color}

\begin{document}
\begin{frontmatter}

\title{Phase space deformations in phantom  cosmology.}

\author{J. L. L\'opez$^a$\corref{cor1}}
\ead{ jlopez87@uabc.edu.mx}
\author{M. Sabido$^{b,c}$\corref{cor2}}
\ead{msabido@fisica.ugto.mx}
\author{C. Yee-Romero$^a$\corref{cor3}}
\ead{carlos.yee@uabc.edu.mx}

\address{ $^a$ Departamento de Matem\'aticas, Facultad de Ciencias\\
 Universidad Aut\'onoma de Baja California, Ensenada, Baja California, M\'exico.\\
 $^b$Department of Theoretical Physics, University of the Basque Country UPV/EHU, P.O. Box 644, 48080 Bilbao, Spain.\\
 $^c$ Departamento  de F\'{\i}sica de la Universidad de Guanajuato,\\
 A.P. E-143, C.P. 37150, Le\'on, Guanajuato, M\'exico.\\
}%
\begin{abstract}
We discuss the physical consequences of  general phase space deformations on the minisuperspace of  phantom cosmology.
Based on the principle of physically equivalent descriptions in the deformed theory, we investigate for what values 
of the deformation parameters the arising descriptions  are physically equivalent. 
We also construct and solve the quantum model and  derive the semiclassical dynamics. 
\end{abstract}

\end{frontmatter}

\section{Introduction}

One of the most intriguing aspects in physics, is the current acceleration of the Universe.  Is it a consequence of modifications to general relativity (GR) or is it a new kind of matter that drives this acceleration?. Although having some theoretical problems, the best answer to this question is the cosmological constant $\Lambda$.  An alternative
 candidate that has been successful  for the description of dark energy is the scalar field  \cite{SF1,ratra,SF2,SF3,Nojiri,Pad}. 
In particular,  scalar fields with a negative kinetic term have been considered in the literature. This type of fields are known as Phantom fields. They have some  interesting properties 
that allows them to be considered as a strange but viable matter which could be relevant in the evolution of the Universe \cite{Rybakov}. In particular, a phantom field  provides an effective negative pressure and a repulsive effect on the matter content of the Universe which in the long term could be responsible for the current accelerated expansion \cite{Perlmutter,Riess1,Riess2,Riess3,Kiefer}.
Therefore it has 
been considered as the matter source of the late time accelerated 
expansion of the Universe \cite{Cald1,Cline,Hu}. 

When considering the spacetime structure of the Universe it is usually done in reference to GR. But when regarding the micro structure of spacetime we do not have a universally accepted quantum theory of gravity. Although there are several candidates, noncommutativity has been considered as an alternative to understand the small scale structure of the Universe and help in the construction of quantum theory of gravity. For this reason,  
noncommutative versions of gravity have been constructed
\cite{Obregon1,Calmet,Wess,NC2}.
Noncommutativity is usually believed to be present near Planck's scale and is consistent with a discrete nature of spacetime. Motivated by this idea, it is {justified to consider an inherently noncommutative spacetime at the early ages of the universe. Directly using noncommutative  gravity is quite difficult, this is a consequence of the highly nonlinear character of these theories\cite{Obregon1,Calmet,Wess,NC2}. Fortunately there is an alternative, in \cite{Obregon2} the authors introduce} the effects of noncommutativity  using the Moyal product of functions on the Wheeler-DeWitt (WDW) equation.\\
The effects  of the noncommutative deformation at the classical level was studied by a WKB approximation of noncommutative quantum model \cite{eri1} and also by  modifying the Poisson algebra \cite{Obregon3,bastos}. More general minisuperspace deformations have been done in connection with $\Lambda$ \cite{NC3,vakili,Miguel1,yee,eri2,Miguel2}. Phase space deformations give rise to two physically nonequivalent descriptions, the {\it``C-frame"} based on the original variables but with a modified interaction and the {\it``NC-frame"} frame based on the deformed variables but with the original interaction. 
Given this ambiguity, in \cite{Miguel2} a principle was proposed to restrict the value of the deformation parameters in order to make both descriptions physically equivalent.
  
In this work we consider a Friedmann-Robertson-Walker (FRW) cosmological model coupled to a phantom scalar field and study the physical consequences 
of introducing general phase space deformations on the minisuperspace of the theory. We will study both, the classical and quantum models and we also find that the semiclassical approximation of the deformed quantum model agrees with the classical model. The deformation  parameter space is determined  by considering the principle of physically equivalent frames \cite{Miguel2}. 

The paper is organized as follows. In section 2, the commutative model is presented. In  section 3, the minisuperspace phase space deformation
is implemented and the dynamics in the two frames is obtained. Also the deformation parameters are constrained by imposing  physical equivalence between the {\it``C-frame"} and the {\it``NC-frame"}. 
 The quantum analysis is done in section 4, we find the solution to the deformed WDW equation 
and fix the parameters in the deformation in order to make the quantization possible. We also show that the classical paths  arise from the 
semiclassical approximation of the WDW equation obtained from the deformed Hamiltonian. Finally, section 5 is 
devoted for concluding remarks.   
         
\section{Phantom field model}

We start with the flat FRW metric

\begin{equation}
ds^2 = -N^2(t)dt^2 + a^2(t)[ dr^2 + r^2d\Omega]
\end{equation}
$a(t)$ corresponds to the scale factor and $N(t)$ is lapse function.
In this background the action of a minimally coupled phantom scalar field $\varphi (t)$ with constant potential is 
\begin{equation}\label{accion}
S = \int dt \left \{\   -\frac{3a \dot{a}^2}{N} - a^3\left(  \frac{\dot{\varphi}^2}{2N} + N\Lambda    \right)    \right \}\ ,
\end{equation}   
where we have set the units so $8\pi G = 1$. The minus sign in the kinetic term of the scalar action is the difference between the usual scalar field and the phantom scalar field \cite{Kiefer}. The canonical Hamiltonian derived from Eq.(\ref{accion}) is

\begin{equation}\label{ham}
-N \left[ \frac{P_a^2}{12a} +\frac{P_{\varphi}^2}{2a^3} - a^3 \Lambda \right]. \label{Ham1}
\end{equation}
With the  the 
following change of variables
\begin{eqnarray}
 x &=& \mu^{-1}a^{3/2}\sin{(\mu \varphi)}, \label{Trans1} \\ \nonumber
 y &=& \mu^{-1}a^{3/2}\cos{(\mu \varphi)},
\end{eqnarray} 
and  $\mu = \sqrt{3/8}$. The Hamiltonian Eq.(\ref{ham}) can be rewritten as a sum of two harmonic oscillators
\begin{equation}
H = N\left(  \frac{1}{2}P_x^2 + \frac{\omega^2}{2}x^2   \right) + N\left(  \frac{1}{2}P_y^2 + \frac{\omega^2}{2}y^2   \right), \label{Ham1}
\end{equation}
where $\omega^2 = -\frac{3}{4}\Lambda$. When one considers the usual scalar field, the Hamiltonian is transformed to a ``ghost oscillator'' which is simply a difference of two harmonic oscillators  \cite{Miguel1,Miguel2}. 
\section{Deformed Phase Space Model}

There exist different approaches to incorporate noncommutativity into physical  theories. Particularly, in cosmology there is a broadly explored path to study noncommutativity \cite{Obregon2},
where noncommutativity is realized  in the so called minisuperspace variables. 
We will follow  a deformed phase space approach. The deformation is introduced by the Moyal bracket 
$\{f,g\}_{\alpha}=f\star_{\alpha}g-g\star_{\alpha}f$. By substituting  the usual product with the Moyal product 
$(f\star{g})(x)=\exp{\left[\frac{1}{2}\alpha^{ab}\partial_{a}^{(1)}\partial_{b}^{(2)}\right]}f(x_{1})g(x_2)\vert_{x_1=x_2=x}$ such that
\begin{eqnarray}
\alpha =
\left( {\begin{array}{cc}
 \theta_{ij} & \delta_{ij}+\sigma_{ij}  \\
- \delta_{ij}-\sigma_{ij} & \beta_{ij}  \\
 \end{array} } \right) .
\end{eqnarray}
The $2\times 2$ antisymmetric matrices $\theta_{ij}$ and $\beta_{ij}$ represent the noncommutativity in the coordinates and momenta respectively. The $\alpha$ deformed algebra becomes 

\begin{equation}
\{x_i,x_j\}_{\alpha}=\theta_{ij},\{x_i,p_j\}_{\alpha}=\delta_{ij}+\sigma_{ij},\; \{p_i,p_j\}_{\alpha}=\beta_{ij}.\label{alg}
\end{equation}
In this work we use the particular deformations, $\theta_{ij}=-\theta\epsilon_{ij}$ and $\beta_{ij}=\beta\epsilon_{ij}$.  

\noindent There is an alternative to derive a similar algebra to Eq.(\ref{alg}). For this we follow the procedure given in  \cite{NC3,Miguel1}. Start with  the  transformation
 \begin{eqnarray}
\widehat{x}=x+\frac{\theta}{2}P_{y}, \qquad \widehat{y}=y-\frac{\theta}{2}P_{x},\nonumber \\
\widehat{P}_{x}=P_{x}-\frac{\beta}{2}y, \qquad \widehat{P}_{y}=P_{y}+\frac{\beta}{2}x, \label{nctrans}
\end{eqnarray}
on the classical phase space variables $\{x,y,P_x,P_y\}$, these are the variables that satisfy the usual Poisson algebra.
The new variables satisfy a deformed algebra
\begin{equation}
\{\widehat{y},\widehat{x}\}=\theta,\; \{\widehat{x},\widehat{P}_{x}\}=\{\widehat{y},\widehat{P}_{y}\}=1+\sigma,\; \{\widehat{P}_y,\widehat{P}_x\}=\beta,\label{dpa}
\end{equation}
where $\sigma=\theta\beta/4$. Furthermore, as in \cite{NC3,Miguel1}, we assume that the deformed variables satisfy the same relations as their commutative counterparts. 
The resulting algebra is the  same, but the Poisson bracket is different in the two algebras. In Eq.(\ref{alg}),  the brackets are the 
$\alpha$ deformed ones related to the  Moyal product, for the other algebra the brackets are the usual Poisson brackets.

To  construct the deformed theory, we start with a Hamiltonian which is formally analogous to Eq.(\ref{Ham1}) but constructed with the variables that obey the modified algebra Eq.(\ref{dpa}), this gives the deformed Hamiltonian

\begin{equation}
H_{nc} = \frac{1}{2} \left[   (P_x^2 + P_y^2) + \ell^2(xP_y-yP_x) + \widetilde{\omega}^2(x^2+y^2)    \right], \label{Ham2}
\end{equation}
where $\ell^2$ and $\widetilde{\omega}^2$ are given by

\begin{equation}
\ell^2 = \frac{\beta + \omega^2 \theta}{1 + \frac{\omega^2 \theta^2}{4}}, ~~~~~\tilde{\omega}^2 = \frac{\omega^2 + \frac{\beta^2}{4}}{1 + \frac{\omega^2 \theta^2}{4}}.
\end{equation}
There is a significant difference between the transformed Hamiltonian of the scalar field cosmology model \cite{Miguel1} and  Eq. (\ref{Ham2}), the deformed Hamiltonian of the phantom cosmology model. 
Unlike the scalar field case, the crossed term involving position and momentum in the phantom model, corresponds to an angular momentum term. To understand the physics of the deformation, it is
necessary to remember that the deformation Eq.(\ref{nctrans}) defines two physical nonequivalent descriptions, the {\it``C-frame"} where the effects of the deformation are interpreted as a commutative 
space $(x,y)$ but with modification of the original interaction and 
the {\it``NC-frame"} where we work with the deformed variables $(\widehat{x},\widehat{y})$ and the original interaction.  
In general, the dynamics in the two frames is different but the physical interpretation in the {\it``C-frame"} is easier. 
In this frame we can interpret the deformed model as a bidimensional harmonic oscillator  
and the minisuperspace deformation comes into play as an angular momentum term in the Hamiltonian. 
For this reason we will do the calculations in the {\it``C-frame"}. 

We obtain the equations of motion from the Hamiltonian Eq.(\ref{Ham2}), which  in the $(x,y)$ variables are given by 

\begin{eqnarray}\label{nceqm}
\dot{x}&=& P_{x}-\frac{1}{2}\ell^2{y},\qquad
\dot{y}= P_{y}+\frac{1}{2}\ell^2{x},\\ 
{\dot{P}_{x}}&=&-\frac{1}{2}\ell^2 P_{y}-\widetilde{\omega}^2{x}, \qquad
{\dot{P}_{y}}=\frac{1}{2}\ell^2 P_{x}-\widetilde{\omega}^2{y}, \nonumber
\end{eqnarray}
and get
\begin{eqnarray}
 \ddot{x} + \ell^2 \dot{y} + \left(  \widetilde{\omega}^2 -\frac{\ell^4}{4}  \right)x &=&0, \label{EqMot} \\ \nonumber
 \ddot{y} - \ell^2 \dot{x} + \left( \widetilde{\omega}^2 -\frac{\ell^4}{4}  \right)y &=& 0.
\end{eqnarray} 
With the transformation $z=x+iy$ we can easily solve Eq.(\ref{EqMot}). We  have three different solutions depending on the sign of $\widetilde{\omega}^2$.
For $\widetilde{\omega}^2 > 0$, we get

\begin{eqnarray}\label{Sol1}
x(t) &=& A_1 \cos{[( \ell^2/2 + |\widetilde{\omega}|)t ]}
 + B_1  \cos{[(\ell^2/2 - |\widetilde{\omega}|)t]},  ~~~~~~\\ 
y(t) &=& A_1 \sin{[(\ell^2/2 +|\widetilde{\omega}|)t]}
 + B_1  \sin{[(\ell^2/2 - |\widetilde{\omega}|)t]},~~~~~~\nonumber
\end{eqnarray}  
$A_1,B_1$ are arbitrary constants that, from the Hamiltonian constraint,  satisfy $(\ell^2+2|\widetilde{\omega}|)A_1-(\ell^2-2|\widetilde{\omega}|)B_1=0$.
When $\widetilde{\omega}^2 < 0$ we have the  solutions
\begin{eqnarray}\label{Sol2}
 x(t) &=&  \left[A_3 \cosh{(|\widetilde{\omega}| t)}+B_3 \sinh{(|\widetilde{\omega}| t)}\right]\cos{\left( \frac{\ell^2}{2} t \right)},  \\ \nonumber
 y(t) &=&  \left[A_3 \cosh{(|\widetilde{\omega}| t)}+B_3 \sinh{(|\widetilde{\omega}| t)}\right]\sin{\left( \frac{\ell^2}{2} t \right)},
\end{eqnarray}  
for this case the integrating constants satisfy the condition $A_3^2=B_3^2$.
The last solution is for the case $\widetilde{\omega}= 0$, 
\begin{eqnarray} \label{Sol3}
x(t) &=& \left(A_2 +B_2 t\right) \cos{ \left(\frac{\ell^2}{2}  t \right)}, \\ 
 y(t) &=& \left(A_2 +B_2 t\right) \sin{ \left(\frac{\ell^2}{2}  t \right)}, \nonumber
\end{eqnarray} 
the constraint  imposes $B_2=0$. The condition arises when the cosmological constant satisfies  $2\beta = -3\Lambda$.

In analyzing the evolution of the Universe, we will focus our attention on its volume and  calculate $a^3(t)$ in the two frames. Let us start in the  {\it``C-frame"}, 
using the solutions for the model, Eq.(\ref{Sol1}), Eq.(\ref{Sol2}) and Eq.(\ref{Sol3}), together with the change of variables in Eq.(\ref{Trans1}) we get 
\begin{equation}
a^3(t) = \left \{ 
\begin{array}{ccc} \label{CVolume}
& V^{(1)}_0 + V^{(1)}_1 \cos^2{(\widetilde{\omega}t)}  , ~~~~~~~~~~~ \widetilde{\omega}^2 > 0, \\
& ~ \\
& V^{(2)}_0 ,  ~~~~~~~~~~~~~~~~~~~~~~~~~~~~~~~ \widetilde{\omega}^2 = 0, \\
& ~ \\
& V^{(3)}_0 e^{2\tilde{\omega} t}, ~~~~~~~~~~~~~~~~~~~~~~~~~ \widetilde{\omega}^2 < 0, \\
\end{array} \right.
\end{equation}
where $V^{(i)}_0,V^{(i)}_1$ are constructed from the integrating constants. For the first case, $V^{(1)}_0=(A_1-B_1)^2$ and $V^{(1)}_1=4A_1B_1$, we must remember that the 
constants are related. For the second case the volume is the square of a constant. Finally for the last case we have $V^{(3)}_0=A_3^2$, but if we take $A_3=-B_3$ we  
have an exponentially decaying  volume. 

To study the dynamics in the {\it ``NC-frame"} we use the {\it ``C-frame''} solutions. The volume is constructed with the {\it ``NC-frame"}  variables, $\widehat{a}^3(t) = m^2(\widehat{x}^2 + \widehat{y}^2)$   
\begin{equation}
\widehat{a}^3(t) = \left\{ 
\begin{array}{ccc} \label{NCVolume}
& \widehat{V}^{(1)}_0 + \widehat{V}^{(1)}_1 \cos^2{(\widetilde{\omega}t)}  ,  ~~~~~~~~~~~ \widetilde{\omega}^2 > 0, \\
& ~ \\
& \widehat{V}^{(2)}_0,  ~~~~~~~~~~~~~~~~~~~~~~~~~~~~~~~~ \widetilde{\omega}^2 = 0, \\
& ~ \\
& \widehat{V}^{(3)}_1e^{2\tilde{\omega} t}, ~~~~~~~~~~~~~~~~~~~~~~~~~~ \widetilde{\omega}^2 < 0, \\
\end{array} \right.
\end{equation}   
where $\widehat{V}^{(i)}_0,\widehat{V}^{(i)}_1$ are constants related to $V^{(i)}_0,V^{(i)}_1$. 
{It is important to mention that even if we have a positive volume in the {\it``C-frame"} we might need to impose some restriction in order to have a valid description in the {\it``NC-frame"}.}
In order to determine the valid range of the deformation parameters  we use the principle used in  \cite{Miguel2} which states the following, {\it
``Deformed phase space models are only valid when the {\it ÒNC-frameÓ} and {\itÒC-frameÓ} descriptions are physically equivalent''}.  
To impose that the description in the two frames should be physically equivalent, we start by analyzing  the expression  for the  {\it ``C-frame"} volume.  
It is easy to show that the volume is always positive. In the first case one obtains that $V^{(1)}_0+V^{(1)}_1=(A_1+B_1)^2\ge0$ so is positive for any $t$. 
For the second case the volume is also positive, and finally for the third case the constant $V^{(3)}_1>0$. 
Then we  conclude that for the three cases,  the  {\it``C-frame"} volume is always positive for any choice of the integration constants.
{The constants in  the {\it``NC-frame"} solutions are related to the {\it``C-frame"} constants as follows. For the first case, $\widehat{A}_1=(1+|\widetilde{\omega}|\theta/2)A_1$ and $\widehat{B}_1=(1-|\widetilde{\omega}|\theta/2)B_1$, for the second case the volume is the same and for the last case the constant in the {\it``NC-frame"} is related by $\widehat{A}^2_3=(1+\tilde{\omega}^2\theta^2/4)A_3^2$.}

\noindent We see from Eq. (\ref{CVolume}) and Eq.(\ref{NCVolume}) that  for $\theta = 0$, the two frames are physically equivalent. We can conclude that $\theta=0$ and $\beta\ne0$ are valid values for the deformation parameters.
Now we consider the general case $\theta \neq 0$ and $\beta\ne0$. It is clear that in the first two cases, that the volume has the same physical behavior in the two frames then there are 
no restriction for the deformation parameters.
For the third case, we have the same behavior as long as we satisfy the condition
$1+\frac{\theta^2 \widetilde{\omega}^2}{4} >0$. 
Furthermore, the evolution is that of an exponential  scale factor for an accelerating Universe. Then comparing with
 a de-Sitter Universe we find (only for the third case), an effective positive 
cosmological constant 

\begin{equation}
\Lambda_{Eff} = \frac{1}{3}\left(  \frac{3\Lambda - \beta^2}{1-\frac{3\theta^2\Lambda}{16}}  \right). \label{LambdaEf}
\end{equation}
This effective cosmological constant is determined by the deformation parameters as well as the value of $\Lambda$. Since current observations point to a positive 
cosmological constant. This gives and extra condition that restricts the values of the deformation parameters  by imposing  that $\Lambda_{eff}>0$. After fixing the value 
of $\Lambda$, we have three definite restrictions: 
\begin{enumerate}
\item[i)] $\Lambda_{Eff} > 0$. \label{Conditions} 
\item[ii)] $f(\beta,\Lambda) = 3\Lambda -\beta^2$  and $g(\theta,\Lambda) = 1-\frac{3\theta^2 \Lambda}{16}$  must have the same sign. \\ \nonumber
\item[iii)] $ 1+\frac{\theta^2 \widetilde{\omega}^{ 2}}{4} > 0$.
\end{enumerate}
From ii), we can see in Fig.(1) and Fig.(2) the regions where these inequalities are satisfied. We did not consider $\Lambda \leq 0$, since condition ii) would not be satisfied, 
for this reason the plots only show the region where  $\Lambda > 0$. 
\begin{figure}[h]
\centering
\includegraphics[width=7cm]{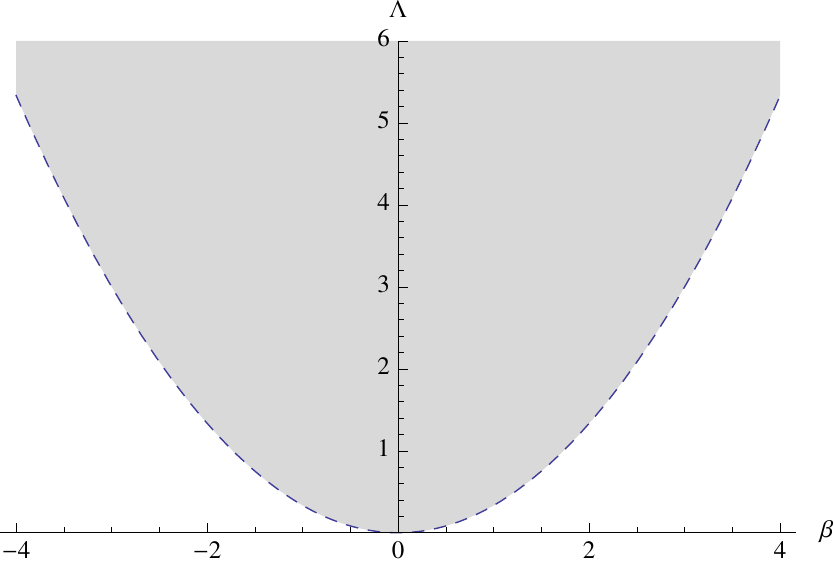}
\label{Figure1}
\caption{This figure shows, in the gray region, the permitted values of $\beta$ and $\Lambda$ that satisfy $f(\beta,\Lambda) > 0$. The white region
corresponds to he values where $f(\beta,\Lambda) < 0$.}
\end{figure}

\begin{figure}[h]\
\centering
\includegraphics[width=7cm]{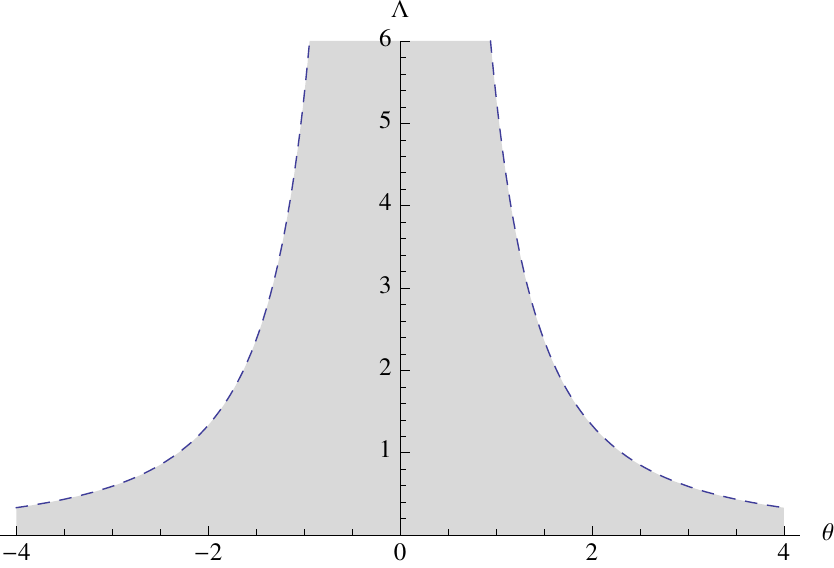}
\label{Figure2}
\caption{The gray region in this figure shows the values of $\theta$ and $\Lambda$ that satisfy $g(\theta,\Lambda) > 0$. The white region
corresponds to the values where $g(\theta,\Lambda) < 0$.}
\end{figure}
\noindent We want to emphasize that for the phantom scalar field we cannot
define an effective cosmological constant when $\Lambda = 0$, as it was done for the usual scalar field in \cite{Miguel1}. When setting $\Lambda=0$ 
we will have $\widetilde{\omega}^2>0$ meaning that we are in the first case and therefore we will not have an exponential behavior. For this model
the deformation maps an initial positive cosmological constant model to another positive cosmological constant model, 
the two cosmological constants have different values but they are both positive. This fact also leads to a more restrictive 
space of parameters than the one analyzed 
in \cite{Miguel2}.  
 
\section{Noncommutative quantum model}

Now we proceed with the quantization of the deformed Hamiltonian. The usual canonical quantization approach $H_{nc} \psi = 0$, gives the corresponding WDW equation. 
The Hamiltonian $H_{nc}$, explicitly contains the angular momentum operator $(xP_y - yP_x) = L_z$
and can be written in terms of the Hamiltonian of a 2-dimensional harmonic oscillator  $H_{xy}$, with frequency $\widetilde{\omega}^2$ plus an angular momentum term, that is 
$H_{nc} = H_{xy} + L_z$. Using the similarity of the WDW equation, with the 2-dimensional harmonic oscillator, we will follow the approach in \cite{quantum,quantum2}. First we write the eigenvalue equation for the harmonic oscillator, then we will find the energy eigenvalues and finally impose the zero energy condition. After writing the quantum equation in  polar coordinates,  
we obtain the equation for the energy eigenstates, and  is given by

\begin{equation}
-\frac{\hbar^2}{2}\left[  \frac{\partial^2 \psi}{\partial r^2} + \frac{1}{r}\frac{\partial \psi}{\partial r} + \frac{1}{r^2}\frac{\partial^2 \psi}{\partial \phi^2}  \right] 
+ \frac{\widetilde{\omega}^2}{2}r^2\psi -\frac{i\hbar \ell^2}{2} \frac{\partial \psi}{\partial \phi} = E \psi. \label{WDW}
\end{equation} 
It is well known that the Hamiltonian of the 2d harmonic oscillator commutes with the angular momentum. This means that we have $[H_{xy}, L_z] = 0$, allowing us to write 
the wave function with eigenstates common to both operators. This breaks the degeneracy 
in energy eigenstates  and every state is uniquely specified by the quantum numbers $n$ and $m$. Solving Eq.(\ref{WDW}) we find  the wave function of the Universe $\psi_{n,m} (r,\phi)$  

\begin{equation}
\psi_{n,m}(r,\phi) = N \left( \frac{\widetilde{\omega}r^2}{\hbar} \right)^{\frac{|m|}{2}} L^{|m|}_n\left( \frac{\widetilde{\omega}r^2}{\hbar} \right)e^{im\phi}e^{-\frac{\tilde{\omega} r^2}{2 \hbar}}. \label{Wave}\end{equation}
where  $n=0,1,2,...\:; ~m = 0, \pm 1, \pm 2,...,\pm n,$ and $L_n^{\alpha} (z)$  are the generalized Laguerre polynomials. 
The zero energy condition 
$E = 0$ from the WDW equation  is trivially satisfied by $\ell = \widetilde{\omega} = 0$. For 
$\ell \neq 0, ~\widetilde{\omega} \neq 0$ (for instance for a single state of quantum numbers $n$ and $m$) we get the general equation  

\begin{equation}
\frac{\ell^2}{\widetilde{\omega}} = -\frac{2(2n + |m| + 1)}{m}. \label{Condition2}
\end{equation}
By using the definitions for $\ell^2$ and $\widetilde{\omega}^2 > 0$ we can write this equation as
 
 \begin{equation}
 \frac{16\beta -12\theta \Lambda}{\sqrt{(16-3\theta^2 \Lambda)(4\beta^2-12\Lambda)}} = -\frac{2(2n + |m| + 1)}{m}. \label{Condition3}
 \end{equation}
With respect to the condition in Eq. (\ref{Condition3}), we need to find the values of $\theta$, $\beta$ and $\Lambda$ to satisfy the condition. 
The first thing to notice is that the right hand side of this equation takes only discrete values and to determine if there are possible 
values of the parameters to satisfy the equation we may proceed as follows. Analyze the behavior 
of the function $h(\theta,\beta,\Lambda) = \frac{16\beta -12\theta \Lambda}{\sqrt{16-3\theta^2 \Lambda}\sqrt{4\beta^2-12\Lambda}}$
and then  see whether or not, there are permitted values of the parameters for which the function takes the value $h(\theta,\beta,\Lambda) = -\frac{2(2n + |m| + 1)}{m}$. 
The quantum solution is possible when $\widetilde{\omega}^2> 0$, which corresponds to the first solution in Eq. (\ref{CVolume}) and Eq.(\ref{NCVolume}). Therefore, the space 
of parameters is already depicted in Fig.(1) and Fig.(2), corresponding to the condition that the functions $f(\beta,\Lambda)$ and $g(\theta,\Lambda)$ have opposite sign.
The valid values of parameters $\theta$, $\beta$ and $\Lambda$, corresponds to the gray region of Fig. 1 and white region of Fig. 2, or vice versa.   
We use the inverse transformation of Eq.(\ref{Trans1}) with Eq. (\ref{Wave})  
\begin{equation}
r^2 = \frac{a^3}{\mu^2}, ~~~~~ \phi =  \frac{\pi}{2}-\mu \varphi,
\end{equation}
to  find the wave function $\psi_{n,m} (a,\varphi)$ in terms of the scale factor $a$ and the phantom field $\varphi$.
 
\noindent We can construct a general wave packet 

\begin{equation}
 \psi(a,\varphi) = \sum_n^N c_{n,m}\psi_{(n,m)} (a,\varphi), \label{WaveP}  
\end{equation}  
 with $m$ fixed and $c_{n,m}$ arbitrary constants as long as the condition $E = \sum_n^N E_{(n,m)} = 0$ also holds. 
 When this is the case, it is always possible to find values of $\theta$, $\beta$ and $\Lambda$ 
to satisfy the constriction Eq. (\ref{Condition2}), under which these values are fixed unambiguously. For a given value of $m$, the sum in Eq.(\ref{WaveP}) runs over odd/even values of $n$
depending on the odd/even value of $m$ respectively. The wave packet is normalizable by the use of 
the orthogonality relation of generalized Laguerre polynomials.
%
The total energy of the wave packet 
%
will satisfy the generalized zero energy condition  Eq.(\ref{Condition2}), but replacing $n$ by $N$.
%
From  the arguments given above, we conclude that it is always possible to find, for a given $\Lambda > 0$  values for $\theta$ and $\beta$ for which the condition holds. 

\subsection{Classical paths from a WKB approximation}

One can apply a WKB type approximation  on the WDW equation $H_{nc} \Psi = 0$, to see if we get the classical solutions. 
First we notice that the effective classical cosmological constant 
given in Eq. (\ref{LambdaEf}) is the same quantum cosmological constant  in the  Hamiltonian $H_{nc}$.
Remembering that in the commutative Hamiltonian  $4\omega^2 = -3\Lambda$, for the deformed Hamiltonian 
$4\widetilde{\omega}^2 = -3\Lambda_{Eff}$. This is the generalized deformed frequency  and using the 
definition of $\widetilde{\omega}^2$ we get the same effective cosmological constant. 

We now apply the 
standard procedure for the semiclassical approximation. We propose a wave function of the form 

\begin{equation}
\psi \propto e^{\frac{i}{\hbar}S_1(x) + \frac{i}{\hbar}S_2(y)},
\end{equation}  
which upon substitution in $H_{nc} \Psi = 0$, in the limit $\hbar \rightarrow 0$ and using the approximation

\begin{equation}
\left( \frac{\partial S_1(x)}{\partial x} \right)^2 > > \frac{\partial^2 S_1(x)}{\partial x^2},~~~ \left( \frac{\partial S_2(y)}{\partial y} \right)^2 > > \frac{\partial^2 S_2(y)}{\partial y^2},
\end{equation}
we get the Einstein-Hamilton-Jacobi (EHJ) equation

\begin{eqnarray}
&& \left( \frac{\partial S_1(x)}{\partial x} \right)^2 + \left( \frac{\partial S_2(y)}{\partial y} \right)^2 + \widetilde{\omega}^2(x^2 + y^2) \\ \nonumber
& +& \ell^2x\left( \frac{\partial S_2(y)}{\partial y} \right) -\ell^2 y \left(  \frac{\partial S_1(x)}{\partial x}  \right)  = 0.
\end{eqnarray}
With the identification $\frac{\partial S_1(x)}{\partial x} = P_x$, $\frac{\partial S_2(y)}{\partial y} = P_y$,  the EHJ equation takes the form

\begin{equation}
\dot{x}^2 +\dot{y}^2 +\left( \widetilde{\omega}^2-\frac{\ell^4}{4} \right) \left(x^2+y^2\right)  = 0.
\end{equation}
Now we take the time derivative of the EHJ equation and divide the result by $\dot{x}\dot{y}$  to get
\begin{equation}
\frac{\ddot{x}}{\dot{y}}+\left( \widetilde{\omega}^2-\frac{\ell^4}{4}\right)\frac{x}{\dot{y}}+\frac{\ddot{y}}{\dot{x}}+\left( \widetilde{\omega}^2-\frac{\ell^4}{4}\right)\frac{y}{\dot{x}}=0,
\end{equation}
we can separate the expression by equating the first two terms to $-\ell^2$ and the last two to $\ell^2$. The two resulting equations are the equations of motion in Eq.(\ref{EqMot}). We can be confident that  
the same classical paths arise in the classical limit of the  WDW equation obtained from the deformed Hamiltonian.   
\section{Conclusions}

In this paper we have studied the effects of the minisuperspace phase space deformations in the dynamics of  phantom scalar cosmology. 
The deformation was applied on the minisuperspace variables. From the deformed Hamiltonian we derived the  equations of motion. It is known that from the deformation two generally nonequivalent descriptions  arise. Using the solutions to Eq.(\ref{EqMot}), we calculate the volume of the Universe in the  {\it``C-frame"} and in the
{\it``NC-frame"}. {To constrain the deformation parameters $\beta$ and $\theta$, we impose that for phase space deformations to be valid, the physical description in the two frames must be equivalent. From the value of $\widetilde{\omega}^2$ three different cases appear.  
For $\widetilde{\omega}^2>0$, we get a bouncing Universe and the solutions in the two frames are equivalent. For the last case, when the Hamiltonian constraint is applied, we have a de Sitter Universe and the parameter $\widetilde{\omega}^2$ can be interpreted as an effective positive cosmological constant $\Lambda_{Eff}$. The value of $\Lambda_{Eff}$  depends on the  deformation parameters $\theta$ and $\beta$ as well as the original $\Lambda$.}
If we set the original cosmological constant $\Lambda = 0$, then $\widetilde{\omega}^2>0$ 
and is not possible to have $\Lambda_{Eff} \geq 0$. 
This imposes a stronger restriction on the values of $\beta$ and $\theta$ 
and the parameter space is even more restricted, but even when this is the case we have a region of values where the two descriptions
are physically equivalent.

We quantized the model, obtaining the corresponding WDW equation and found the wave function of the Universe. We derived a relation that fixes the values of the parameters 
$\theta$, $\beta$ and $\Lambda$ in order to satisfy the zero energy condition of the WDW equation. We showed that the deformed phantom cosmological model is physically viable. Also, when $\widetilde{\omega}^2 > 0$ the quantization can be achieved and the introduction of the deformation breaks the degeneracy of the quantum states. 
A general condition on the parameters $\theta$ and $\beta$ arises from the zero energy condition that can be satisfied by the general wave packets. 
{It is worth to mention that there is a nice physical interpretation in the {\it``C-frame"}. It is related to the introduction to an effective perpendicular and constant ``magnetic field" $\vec{B}$. 
This can be seen by taking the bidimensional harmonic oscillator coupled to the vector potential $\vec A=(\frac{B}{2}y,-\frac{B}{2} x)$ where the magnitude of the field will be $B=\ell^2$. 
Then we can see that the zero energy condition is related to the manipulation 
of this field and corresponds to a modification in the deformation parameters $(\theta,\beta)$.} 
Finally, we performed the WKB approximation  
and show that the same classical solutions arise in the classical limit of the WDW equation of the corresponding 
deformed Hamiltonian. 


\section*{Acknowledgments}
This work is supported  by CONACyT  research grants 167335, 257919, 258982.  

\noindent J. L. L\'opez is supported by  PRODEP postdoctoral grant 17-1947. This work is part of the PRODEP research network ``Gravitaci\'on y F\'isica Matem\'atica".

\noindent M. S. is supported by  DAIP1107/2016 and by the CONACyT program  ``Estancias sab\'aticas en el extranjero'',   grant 31065. 



\end{document}